\documentclass[prl,twocolumn,superscriptaddress,showpacs,floats,floatfix]{revtex4}

\usepackage{epsfig}
\begin{document}

\draft

\title{Electric Conductance of Rh Atomic Contacts under Electrochemical Potential 
Control}

\author{Tatsuya Konishi}
\affiliation{Division of Chemistry, Graduate School of Science, Hokkaido University, N10W8, Kita, Sapporo, 060-0810, Japan}

\author{Manabu Kiguchi}
\thanks{Present address: Department of Chemistry, Graduate School of Science and Engineering, 
Tokyo Institute of Technology 2-12-1 W4-10 Ookayama, Meguro-ku, Tokyo 152-8551, Japan, E-mail: kiguti@chem.titech.ac.jp}
 \affiliation{Division of Chemistry, Graduate School of Science, Hokkaido University, N10W8, Kita, Sapporo, 060-0810, Japan}
 \affiliation{PRESTO, Japan Science and Technology Agency, N10W8, Kita, Sapporo, 060-0810, Japan}

\author{Kei Murakoshi}
\affiliation{Division of Chemistry, Graduate School of Science, Hokkaido University, N10W8, Kita, Sapporo, 060-0810, Japan}

\date{\today}

\begin{abstract}
The electric conductance of Rh atomic contacts was investigated under the electrochemical 
potential control. The conductance histogram of Rh atomic contacts varied with the electrochemical 
potential. When the electrochemical potential of the contact was kept at $\Phi_{0}$= 0.1 V vs. Ag/AgCl (Rh 
potential), the conductance histogram did not show any features. At $\Phi_{0}$= -0.1 V (under potential 
deposited hydrogen potential), the conductance histogram showed a feature around 2.3 $G_{0}$ ($G_{0}$ =2$e^{2}/h$), 
which agreed with the conductance value of a clean Rh atomic contact, which was observed in 
ultrahigh vacuum at low temperature. At $\Phi_{0}$= -0.25 V (over potential deposited hydrogen potential), 
the conductance histogram showed features around 0.3 and 1.0 $G_{0}$. The conductance behavior of the 
Rh atomic contact was discussed by comparing previously reported results of other metals, Au, Ag, 
Cu, Pt, Pd, Ni, Co, and Fe. The conductance behavior of the metal atomic contacts related with the 
strength of the interaction between hydrogen and metal surface. 

\end{abstract}

\maketitle

\section{INTRODUCTION}
\label{sec1}
A metal nano contact on an atomic scale is of great interest in both fundamental science and 
nano technology \cite{1}. The electrical conductance through a metal nano contact is expressed by 
$G=2 e ^{2} /h \sum T _{i}$ where $T_{i}$ is the transmission probability of the $i$-th conductance channel, $e$ is the 
electron charge, and $h$ is Plank's constant. Conductance of the metal nano contacts depends not only 
on the atomic structure of the nano contacts but also on inherent properties of the metal. The 
conductance of atomic contacts is $G_{0}$ (2$e^{2}/h$) for noble metals, Au, Ag, Cu, and 1.5-3 $G_{0}$ for 
transition metals, Fe, Co, Ni, Pt, Pd \cite{1}. Various metal nano contacts have been fabricated using a 
scanning tunneling microscope (STM) and mechanically controllable break junctions (MCBJ) 
technique, and their electrical conductance was investigated in air, ultra high vacuum (UHV), and 
solution \cite{2,3,4,5,6,7,8,9,10,11,12}. In the case of Au, the atomic contact can be easily prepared, and conductance 
quantization behavior has been observed even in air and solution at room temperature \cite{4,6,7}. On 
the other hand, it is difficult to prepare atomic contacts showing fixed conductance values for 
transition metals. Featureless conductance histograms were observed at room temperature \cite{1}. The 
difficulty may originate from the chemical and/or mechanical instability of the transition metal 
nano contacts. Previous report revealed that the conductance quantization behavior of the transition 
metal nano contact was sensitive to the molecular adsorption on the surface of the metal nano 
contacts \cite{2}. 

To overcome the difficulty in preparing the stable transition metal nano contacts, the 
electrochemical method has been recognized as a powerful approach \cite{4,5,6,7,8,9,10,11,12}. The electrochemical 
potential determines the potential energy of electrons of the nano contact, resulting in the control of 
the bonding strength between the metal atoms, and the interaction of the metals with molecules of 
surrounding medium. These facts could make possible to fabricate stable metal nano contacts, 
which can not be stabilized in other environment. Using electrochemical method, stable Fe, Co, Ni 
Pt, and Pd atomic contacts have been prepared at room temperature under the hydrogen evolution 
reaction \cite{8,9,10,11,12}. Here, it should be noticed that the conductance of the metal atomic contact 
depended on the species of metals. For example, the conductance histogram of Ni contacts showed 
a feature around 1-1.5 $G_{0}$, which agreed with that obtained for clean Ni contacts prepared in UHV 
at 4 K \cite{2,10}. On the other hand, the conductance histogram of Pd contacts showed a feature around 
1 $G_{0}$, which agreed with that obtained for the Pd contacts in hydrogen environment at 4 K \cite{3,11}. 
The conductance behavior of metal atomic contacts prepared in solution may reflect the strength of 
the interaction between hydrogen and metals. It is important to clarify the relationship between the 
conductance of the metal atomic contacts and the interaction between hydrogen and metal atomic 
contacts. In the present study, we have investigated the conductance of the Rh atomic contacts under 
the electrochemical potential control. The conductance of Rh atomic contacts was discussed by 
comparing that of other metal atomic contacts in order to clarify the relationship between conductance 
of the metal atomic contacts and the interaction between hydrogen and metal atomic contacts.

\section{EXPERIMENTAL}
\label{sec2}

The experiments have been performed with the modified STM (Pico-SPM, Molecular 
Imaging) with the Nano ScopeIIIa controller (Digital Instruments). The detail of the experimental 
setup and condition was described in our previous reports \cite{6,7,8,9,10,11,12}. The STM tip was made of a Au 
wire (diameter $\sim$0.25 mm, $>$99 $\%$) coated with wax to eliminate ionic conduction. A 0.50 mm 
diameter Pt wire was used as a counter electrode. The substrate of Au(111) was prepared by a flame 
annealing and quenching method \cite{13}. The electrochemical potential ($\Phi_{0}$) of the STM tip and 
substrate was controlled using a potentiostat (Pico-Stat, Molecular Imaging) with a Ag/AgCl 
reference electrode. The electrolyte solution was 1mM RhCl$_{3}$, 2mM HCl, 50mM H$_{2}$SO$_{4}$ (Rh 
solution). The Rh atomic contacts were prepared in the following manner. First, the electrochemical 
potential of both the STM Au tip and Au substrate were maintained at values lower than the 
potential where bulk metal deposition proceeded ($\Phi_{0}$= -0.1 V)\cite{14}. After sufficient deposition of Rh with 
estimated thickness of more than a few $\mu m$ both onto the Au surfaces of the STM tip and the 
substrate, the tip was pressed into the substrate and then pulled out from the substrate. Separation of 
these contacts resulted in the formation of Rh atomic contacts between the tip and substrate. 
Conductance was measured during the breaking process under an applied bias of 20 mV between 
the tip and substrate. Statistical data was obtained from a large number (over 3000 times) of 
individual conductance traces.

\section{RESULTS}
\label{sec3}
\begin{figure}
\begin{center}
\leavevmode\epsfxsize=80mm \epsfbox{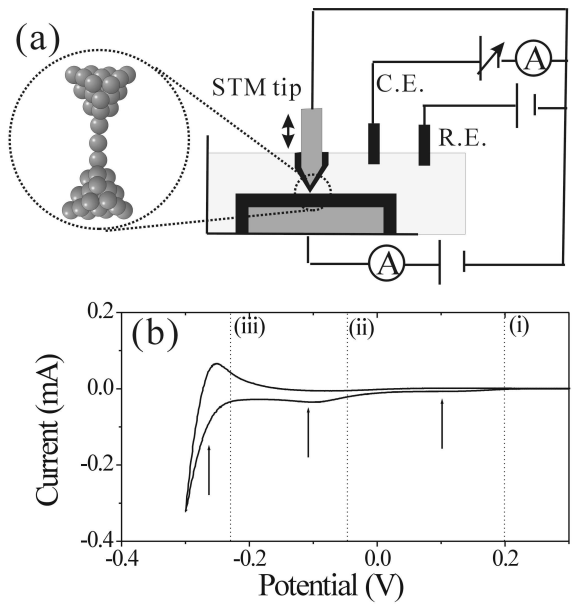}
\caption{(a) Schematic view of the electrochemical STM, C.E. : counter electrode, R.E. : reference 
electrode. (b) Cyclic voltammogram of the Au electrode in the Rh solution. (i), (ii), (iii) are the 
boundary between the Rh deposition, UPD of hydrogen (UPD-H), and hydrogen evolution reaction 
(OPD-H) potential. The arrows indicate the potential measured in the present study.} \label{fig1}
\end{center}
\end{figure}

Figure~\ref{fig1} shows the cyclic voltammogram of the Au electrode in the Rh solution. Bulk Rh 
deposition proceeded at $\Phi_{0}$= 0.2 V. A reduction current due to the under potential deposited 
hydrogen was observed at a potential regime from $\Phi_{0}$= 0.2 V to -0.05 V. The hydrogen evolution 
reaction occurred at a potential more negative than $\Phi_{0}$= -0.2 V \cite{14}. Conductance measurements were 
performed at $\Phi_{0}$= 0.3V (Au potential), $\Phi_{0}$= 0.1 V (Rh potential), $\Phi_{0}$= -0.1 V (UPD-H potential), and 
$\Phi_{0}$= -0.25 V (OPD-H potential).

\begin{figure}
\begin{center}
\leavevmode\epsfxsize=80mm \epsfbox{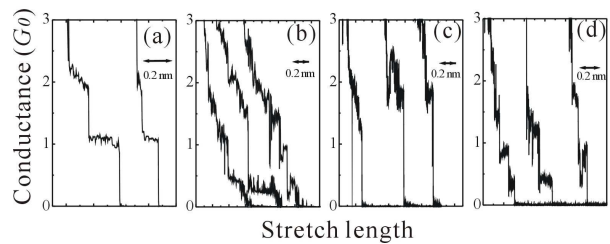}
\caption{
Conductance traces in the Rh solution at (a) $\Phi_{0}$= +0.3 V (Au potential), (b) +0.1 V (Rh 
potential),  (c) -0.1 V (UPD-H potential) and (d) -0.25 V (OPD-H potential).} \label{fig2}
\end{center}
\end{figure}

\begin{figure}
\begin{center}
\leavevmode\epsfxsize=80mm \epsfbox{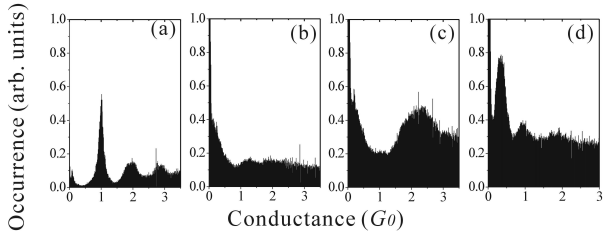}
\caption{
Conductance histograms in the Rh solution at (a) $\Phi_{0}$= +0.3 V (Au potential), (b) +0.1 V (Rh 
potential),  (c) -0.1 V (UPD-H potential) and (d) -0.25 V (OPD-H potential). The conductance 
histograms were obtained from 1000 conductance traces of breaking the contacts. The intensity of 
the conductance histograms was normalized with the number of the conductance traces.} \label{fig3}
\end{center}
\end{figure}

Figures~\ref{fig2} and~\ref{fig3} show conductance traces and conductance histograms of metal contacts in the 
Rh solution obtained at four different electrochemical potential. At the Au potential, conductance 
decreased in a stepwise fashion, with each step occurring at integer multiples of $G_{0}$. The 
corresponding conductance histogram (Fig.~\ref{fig3}(a)) showed a peak at 1.0 $G_{0}$, which corresponded to a 
clean Au atomic contact \cite{1}. At the Rh potential, the conductance trace decreased showing steps. 
The conductance values of the steps varied with the conductance traces, which leaded to the 
featureless conductance histogram (Fig.~\ref{fig3}(b)). At the UPD-H potential, the conductance trace 
frequently showed steps around 2 $G_{0}$. The corresponding conductance histogram (Fig.~\ref{fig3}(c)) showed 
a broad feature around 2.3 $G_{0}$. At the OPD-H potential, the conductance trace showed steps around 
0.3 and 1.0 $G_{0}$. The corresponding conductance histogram (Fig.~\ref{fig3}(d)) showed features around 0.3 and 
1.0 $G_{0}$. The metal contacts in the Rh solution showed four different conductance behaviors depending 
on the electrochemical potential; 1.0 $G_{0}$ (Au potential), featureless (Rh potential), 2.3 $G_{0}$ (UPD-H 
potential) and 0.3 and 1.0 $G_{0}$ (OPD-H potential). 

\section{DISCUSSION}
\label{sec4}
The structure of the Rh atomic contact is discussed based on the present and previous 
experimental results, and theoretical calculation result \cite{8,16,17,18,19}. The hydrogen adsorption and hydrogen 
evolution reaction on a flat Rh electrode has been investigated \cite{16}. At the UPD-H potential, under 
potential deposited (UPD) hydrogen atoms adsorb in the hollow site of the Rh surface. The UPD 
hydrogen is observed only for Pt, Rh, Pd, Ir, and Ru. UPD hydrogen atoms are not observed for Au, Ag, 
Cu, Iron group metals. At the OPD-H potential, over potential deposition (OPD) hydrogen atoms 
adsorb in the atop site of the Rh surface. Both OPD and UPD hydrogen atoms adsorb on the Rh 
surface at the OPD-H potential. Only the OPD hydrogen atoms contribute to the hydrogen evolution 
reaction. The hydrogen evolution proceeds through the following process. First, protons adsorb onto 
the Rh surface (1), and H$_{2}$ (gas) desorbs via surface diffusion and recombination of two adsorbed H 
atoms (2)

Rh+H$_{3}$O$^{+}$ + e $\to$ Rh-H +H$_{2}$O  (1)

2Rh-H$\to$2Rh+H$_{2}$  (2)

Here, it should be noticed that the strength of the metal-hydrogen bond is larger for the UPD 
hydrogen than that for the OPD hydrogen. 

The structure of the Rh atomic contact is discussed based on the above discussion. At the Rh 
potential, we did not observe any features in the conductance histogram, indicating that any preferential 
atomic configurations were not stabilized in solution. In solution, the fluctuation of atoms is large at 
room temperature. The stable atomic contact could not be formed because of the contribution from 
thermally excited motion of solvent and electrolytes molecules surrounding the atomic contacts. 
Therefore, any preferential atomic configurations would not be stabilized at the Rh potential. At the 
UPD-H potential regime, we observed a 2.3 $G_{0}$  feature in the conductance histogram, which agreed with the 
theoretical calculation results of the clean Rh atomic contact \cite{1}. The close agreement between the 
measured and calculated conductance values indicated that the clean Rh atomic contact would be 
stabilized in solution at UPD-H potential. At the UPD-H potential, the UPD hydrogen atoms adsorb 
on the Rh surface. The adsorbed atomic hydrogen could stabilize the Rh atomic contacts. Similar 
stabilization of metal atomic contacts was observed for the Iron group metals under the hydrogen 
evolution reaction \cite{8,10}. At the OPD-H potential, we observed features around 0.3 and 1.0 $G_{0}$
 in  the conductance histogram. Since the conductance values of 0.3 and 1.0 $G_{0}$ were different from the conductance value 
of clean Rh atomic contacts, the 0.3 and 1.0 $G_{0}$ features would not originate from the Rh atomic 
contact. The origin of the 0.3 and 1.0 $G_{0}$ features is discussed based on the previously reported results 
of Pt atomic contacts in hydrogen environment at 4 K, in which similar conductance histogram was 
observed \cite{17,18,19}. The shot noise measurements, vibration spectra of single molecular junctions, 
and theoretical calculation revealed that a single hydrogen molecule bridged between Pt electrodes, 
showing the conductance value of 1.0 $G_{0}$. When additional hydrogen molecules adsorbed on the stem 
part of the single hydrogen molecule junction, the conductance decreased to 0.6 and 0.2 $G_{0}$, 
depending on the atomic configuration of the molecule junction. The 0.5 and 1.0 $G_{0}$ features were 
also observed for the Pt contacts at the OPD-H potential \cite{8}. The 0.5 and 1.0 $G_{0}$ features were 
attributed to the hydrogen molecule bridge. Since the hydrogen evolution process and electronic 
structure of Rh are similar to those of Pt, the 0.3 and 1 $G_{0}$ features 
observed in the present study would originate from the hydrogen single molecule junctions.
In the present system, the atomic configuration and the conductance were well-defined by the controlling 
the electrochemical potential of electrons in the contact at room temperature. 
Additional characteristic of the present electrochemical system is the precise control of the adsorbed species, 
which are not able to achieve in other systems, such as UHV and air. 
The origin of the 2.3, 0.3 and 1.0 $G_{0}$  features observed in the conductance histogram of the Rh contacts could 
be precisely clarified by the theoretical calculations of the electronic structure of the contact and the experimental 
verification of shot noise and vibration spectroscopy measurements of the single molecular 
junctions having the comparable structure.

\begin{figure}
\begin{center}
\leavevmode\epsfxsize=80mm \epsfbox{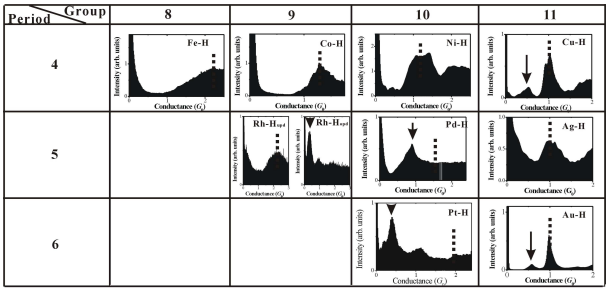}
\caption{
(a) Summary of the conductance histograms of Fe, Co, Ni, Rh, Pd, Cu, Ag, Au, and Pt 
atomic contacts under the hydrogen evolution reaction. The dotted line is the 
conductance value of clean metal atomic contact. 
The arrow shows a feature which is not observed for the clean metal atomic contacts in UHV.} \label{fig4}
\end{center}
\end{figure}

Next, the conductance behavior of Rh contacts is compared with that of other metal contacts. 
Figure~\ref{fig4}(a) shows our previous experimental results of the 
conductance histograms of Fe, Co, Ni, Cu, Pd, Ag, Pt and Au atomic 
contacts under the hydrogen evolution reaction \cite{6,7,8,9,10,11,12}, together with the present Rh results. 
For Rh contacts, the conductance histogram 
at the UPD-H potential is also shown. The conductance histogram roughly depended on the group 
of metal (Gold-Silver-Copper group, Iron group, Platinum group). In the case of Au and Cu contacts, 
a 0.5$\sim$0.6 $G_{0}$ peak appeared in the conductance histogram together with the 1.0 $G_{0}$ peak, which 
corresponded to clean Au and Cu atomic contacts \cite{6,7,12}. The additional 0.5$\sim$0.6 $G_{0}$ peak was not 
observed for the Ag contacts. For Iron group metal contacts, the conductance histogram showed 
features around 1.5$\sim$2.0 $G_{0}$, which agreed with those obtained in UHV at 4 K \cite{9}. For the Platinum 
group metal contacts, the conductance histogram showed features around 0.5 and 1.0 $G_{0}$ \cite{8}. In the case 
of Pt and Pd contacts, the conductance histogram did not show any features at UPD-H potential, while 
conductance histogram of Rh contacts showed a 2.3 $G_{0}$ feature at UPD-H potential. Previous 
experimental and theoretical study proposed the following structure model of the metal atomic 
contacts under the hydrogen evolution reaction. For Au and Cu contacts, hydrogen atoms would 
adsorb on the metal atomic contacts, leading to the appearance of 0.5$\sim$0.6 $G_{0}$ peaks in the 
conductance histogram. For Iron group metals, clean metal atomic contacts would be formed under 
the hydrogen evolution reaction. For Platinum group metals (Pt and Pd), a single hydrogen molecule 
junction would be formed under the hydrogen evolution reaction. Our experimental results 
suggested the formation of the Rh atomic contact and the single hydrogen molecule junction at the 
UPD-H and OPD-H potential, respectively. The Rh atomic contact showed the conductance 
behavior of both Iron group and Platinum group metals, depending on the electrochemical potential. 

The conductance behavior of the metal atomic contacts under the hydrogen evolution reaction 
is discussed by considering the interaction between hydrogen and metals. The free energy of 
hydrogen adsorption on metal surface ($\Delta G$) was investigated by DFT calculation \cite{20}. $\Delta G$ was a 
reasonable descriptor of hydrogen evolution activity for a wide variety of metals \cite{20}. The 
interaction between hydrogen and metal surface decreased with $\Delta G$. The values of $\Delta G$ was 0.66, 
0.28, 0.62, -0.16, -0.13, -0.08, -0.12 eV for Au, Cu, Ag, Co, Ni, Pt, Rh, respectively. The interaction 
between hydrogen and metal surface decreased in the order of Iron group, Platinum group, 
Gold-Silver-Copper group metals. Figure~\ref{fig4} suggests that the clean metal atomic contacts would be 
formed for the Iron group metals, while the hydrogen incorporated or adsorbed metal contacts 
would be formed for the Pt group metals, Au and Cu. It seems that a clean metal atomic contact 
would be formed, when the interaction between hydrogen and metal is large. Hydrogen 
incorporated (hydrogen molecule junction) or adsorbed metal contacts would be formed, when the 
interaction between hydrogen and metal is small. This peculiar conclusion was supported by the 
conductance behavior of Rh atomic contacts observed in the present study. The Rh atomic contact and hydrogen molecule 
junction were formed at the UPD-H and OPD-H potential, respectively. The interaction between 
hydrogen and Rh surface is larger for UPD hydrogen than OPD hydrogen. The strongly interacting 
UPD hydrogen would stabilize the clean Rh atomic contact, while weakly interacting OPD hydrogen 
would lead to the formation of the hydrogen incorporated atomic contact. Here, it should be noticed 
that the above discussion is based on the investigation of hydrogen on flat metal electrodes. The 
interaction between hydrogen and metal atomic contacts would be different from that on the metal 
flat surface. For example, there is little experimental results which directly show the existence of 
hydrogen on the Au surface under the hydrogen evolution reaction. On the other hand, the 
experimental results clearly showed the interaction between hydrogen and Au atomic contact as a 
change of conductance \cite{6,7}. These results indicates the difference between the flat surface and the 
atomic contact. Further investigation is needed to discuss the interaction between hydrogen and the 
metal atomic contact under the hydrogen evolution reaction.

\section{CONCLUSIONS}
\label{sec5}

We have investigated the electric conductance of Rh atomic contacts under the electrochemical 
potential control. The conductance histograms showed no feature, 2.3 $G_{0}$ feature, 0.3 and 1.0 $G_{0}$ 
features at the Rh, UPD-H, and OPD-H potential. Clean Rh atomic contact would be formed at the 
UPD-H potential, while a single hydrogen molecule junction would be formed at OPD-H potential. 
The conductance behavior of the Rh atomic contacts and other metal contacts could be explained by 
the strength of the interaction between hydrogen and metal surface.

\section{ACKNOWLEDGMENTS}
This work was supported by a Grant-in-Aid for Scientific Research on Priority Areas "Electron 
transport through a linked molecule in nano-scale", Effective Utilization of Elements 
"Nano-Hybridized Precious-Metal-Free Catalysts for Chemical Energy Conversion", and the Global 
COE Program (No. B01) from MEXT.

\newpage

\end{document}